\documentclass{aa}

\newif\ifshowlinenumbers
\showlinenumbersfalse

\bibpunct{(}{)}{;}{a}{}{,}

\usepackage{subcaption}
\usepackage{lscape}
\usepackage{placeins}
\usepackage{multirow}

\usepackage[pdftex,dvipsnames]{xcolor}

\usepackage{definitionsII-private}
\usepackage{graphicx}

\usepackage{lipsum}

\usepackage{txfonts}

\usepackage{lipsum}
\usepackage{xargs}
\usepackage{soul}

\usepackage{hyperref}
\usepackage[nameinlink]{cleveref}
\usepackage[normalem]{ulem}
\usepackage{float}
\usepackage{placeins}
\usepackage{bm}
\usepackage{academicons}
\usepackage{natbib}
\usepackage{adjustbox}

\usepackage{hyperref}
\hypersetup{
    colorlinks=true,
    linkcolor=blue,
    filecolor=blue,      
    urlcolor=blue,
    citecolor=blue,
    }

\usepackage{natbib}
\bibpunct{(}{)}{;}{a}{}{,}
\usepackage{xcolor}

\newcommand{\referee}[1]{{ #1}}

\newcommand{\MyDeprecated}[1]{{\color{Red} \st{[DEPRECATED]}   }}

\begin{document}

\title{
Validity of the CRD limit for modeling scattering polarization\\
in the photospheric \SrILA{} line
}
        \authorrunning{Riva et al.}
        \author{Simone Riva \inst{1,2}  \corrauth{simone.riva@irsol.usi.ch}
        \and
        Gioele Janett\inst{1,2} \email{gioele.janett@irsol.usi.ch}
        \and
        Franziska Zeuner\inst{1} \email{franziska.zeuner@irsol.usi.ch}
        \and
        Luca Belluzzi\inst{1,2} \email{luca.belluzzi@irsol.usi.ch}
        \and
        Fabio Riva\inst{1} \email{fabio.riva@irsol.usi.ch}        \and
        Pietro Benedusi\inst{1,2} \email{pietro.benedusi@irsol.usi.ch}
}

\institute{
       Istituto ricerche solari Aldo e Cele Daccò (IRSOL), Faculty of Informatics, Università della Svizzera italiana, CH-6605 Locarno, Switzerland
        \and
        Euler Institute, Faculty of Informatics, Università della Svizzera italiana,
        CH-6962 Lugano, Switzerland
}

\abstract
{
Scattering polarization in the \SrILA{} line has become a key diagnostic tool for the small-scale unresolved magnetic fields in the quiet solar photosphere, which are inaccessible to Zeeman-based techniques.
The  limit of complete frequency redistribution (CRD) is commonly used to model the \SrILA{} line signals, but its validity and impact on the magnetic sensitivity of this line have not yet been systematically investigated.
The proposed study is timely, given the advent of new facilities capable of observing this line with unprecedented accuracy, and for future synoptic programs targeting long-term variations of the quiet-Sun magnetism.
}
{
We aim to systematically assess the CRD limit  against the general partial frequency redistribution (PRD) description of scattering processes for modeling the \SrILA{} scattering polarization, focusing on its magnetic sensitivity through the Hanle effect. 
}
{
We solved the radiative transfer (RT) problem for polarized radiation out of local thermodynamic equilibrium (non-LTE) in a {semi-empirical} one-dimensional plane-parallel \referee{static} atmosphere, considering both the CRD and PRD scattering descriptions, across a range of magnetic field configurations relevant for Hanle diagnostics.
}
{
In the presence of small-scale unresolved magnetic fields, where the magnetic field manifests only through Hanle depolarization of the $Q/I$ signal, we found no significant differences between the CRD and PRD emergent profiles, which also exhibit essentially identical Hanle magnetic sensitivity.
In the presence of deterministic magnetic fields, the CRD approximation remains accurate when the scattering polarization signals are sufficiently strong (typically greater than 0.3\%). For weaker signals, PRD effects may  have an appreciable impact on the emergent polarization profiles.
}
{
We support the use of the CRD description for modeling \SrILA{} scattering polarization in the vast majority of observationally relevant cases, with PRD effects becoming important only for weak polarization signals.
}


\maketitle
\ifshowlinenumbers\else\nolinenumbers\fi

\FloatBarrier
\section{Introduction}
Accurately diagnosing the small-scale 
magnetic fields of the quiet solar photosphere
remains a major challenge in modern solar physics
\citep{mackay2012,delacruz_rodriguez2017}.
Such magnetic fields are  difficult to reveal through Zeeman-based diagnostics because the Zeeman effect is largely insensitive to fields with mixed polarities at spatial scales below the observational resolution due to cancellation effects in the polarization signals \citep[e.g.,][]{Bellot2019}.

A way to probe these unresolved, tangled photospheric magnetic fields is by exploiting the depolarization that they induce, via the Hanle effect, on the linear polarization signals produced in atomic and molecular spectral lines by the scattering of anisotropic radiation \citep[][]{stenflo1982,trujillo_bueno2006}.
{
The \SrILA{} line exhibits a strong scattering polarization signal \citep[e.g.,][]{gandorfer2002} and is sensitive to the Hanle effect in the range between 5\,G and 125\,G. For these reasons, it has been widely exploited to characterize the small-scale magnetic fields of the solar photosphere \citep[e.g.,][]{Faurobert-Scholl1993,stenflo1997sss,trujillo2004}, and it stands as a promising diagnostic tool for investigating their temporal evolution throughout the solar cycle \citep{Bianda2014,Rempel2020,Rast2021}.
}
Moreover, \citet{zeuner2022} recently showed that the \SrILA{} line also carries diagnostic information on small-scale, spatially-structured
magnetic fields through Hanle-induced polarization rotation, consistent with tentative observational signatures reported by \cite{bianda2018}.

In practice, inferring quantitative information on the magnetism of the quiet photosphere via Hanle diagnostics relies on accurate modeling of the scattering polarization signal of \SrILA{}.
This task is computationally challenging, as it implies solving the radiative transfer (RT) problem for polarized radiation out of local thermodynamic equilibrium (non-LTE) in comprehensive 3D
atmospheric models \citep[][]{TrujilloBueno2007,delpinoaleman2018,delpinoaleman2021}.
Moreover, an appropriate description of scattering processes is needed.
In this respect, the limit of complete frequency redistribution (CRD) has always been considered well justified for modeling the \SrILA{} line, noticing that it is a rather weak line in the solar spectrum, with a sharp Doppler core and no wings.
In particular, 
\citet{Faurobert-Scholl1993} assessed the validity of the CRD limit through RT calculations that also consider partial frequency redistribution (PRD) effects under the angle-averaged (AA) approximation.
A systematic assessment of the impact of the CRD approximation on the magnetic sensitivity of the scattering polarization signal of \SrILA{} was, however, beyond the scope of that work. 
Moreover, recent studies on stronger resonance lines, such as Ca~{\sc i} 4227\,{\AA}, have demonstrated that the AA approximation introduces significant inaccuracies in the line-core scattering polarization profiles \citep[e.g.,][]{sampoorna2017,janett2021a,belluzzi2024accurate,benedusi2026}.

Although small, the possible impact of PRD effects on the scattering polarization signal of \SrILA{} might be relevant for interpreting the observations of unprecedented accuracy and sensitivity that the new generation of large-aperture solar telescopes, such as the operational DKIST \citep{rimmele2022} and the future EST \citep{quniteronoda2022}, will provide.
Notably, \citet{Zeuner2025,zeuner2026} recently reported the first 2D spatial maps showing sub-arcsecond scattering polarization structures in the \SrILA{} line in quiet-Sun regions, obtained using ViSP \citep{dewijn_2022_visp} at DKIST.
Assessing the robustness of the CRD approximation for magnetic diagnostics via the \SrILA{} line is also particularly relevant for future synoptic observations aimed at constraining long-term variations of the quiet-Sun magnetism.
In this context, \cite{Rempel2020} suggested that the quiet-Sun small-scale magnetic field strength may only vary by about 7\,G over the solar cycle.
To achieve such diagnostic precision, even small systematic differences introduced by modeling assumptions may be relevant.

For all these reasons
a systematic and quantitative analysis of the impact of the CRD limit on the magnetic sensitivity of \SrILA{} is desirable and timely.
In this article, we thus compare the results of RT calculations carried out in 1D plane-parallel \referee{static} atmospheric models considering the limit of CRD and the PRD case, both under the AA approximation and in its most general angle-dependent (AD) formulation, across a range of magnetic field strengths and configurations relevant for Hanle diagnostics.
The paper is organized as follows.
In Sect.~\ref{sec:rt_problem}, we lay out the non-LTE RT problem for polarized radiation, including scattering polarization.
In Sect.~\ref{sec:methods}, we present the adopted solution strategy and the numerical setting.
In Sect.~\ref{sec:results}, we analyze the synthetic emergent Stokes profiles of the \SrILA{}, comparing calculations with CRD and PRD scattering descriptions.
In Sect.~\ref{sec:conclusions}, we provide final considerations.

\FloatBarrier
\section{Formulation of the problem}
\label{sec:rt_problem}

To model the scattering polarization signal of the \SrILA{} line, we solved the non-LTE RT problem for polarized radiation in 1D models of the solar atmosphere.
\referee{We consider three different descriptions of scattering processes of increasing complexity: the limit of CRD, and including PRD effects, both under the AA approximation and in the most general AD case.}

\subsection{\SrILA{} line}
\label{sec:SrIline}

The \SrILA{} line is induced by the transition
between the ground level of neutral strontium, $5s^2 \; {}^1\rm{S}_0$ and its excited level $5s5p \; {}^1\rm{P}^o_1$.
\referee{We note that both levels are singlets and that the upper one is not radiatively connected to lower-energy levels other than the ground level.
Consequently, the scattering polarization signal of Sr~{\sc i} 4607\,{\AA} can be quantitatively modeled with a simple two-level atom \citep[e.g.,][]{delpinoaleman2021}.}
Having total angular momentum $J=0$, the lower level cannot carry atomic polarization by definition.
Moreover, being it the ground level and thus having a much longer lifetime than the upper level, it can be considered infinitely-sharp.

In spite of being a resonance line, with a relatively high value of the Einstein coefficient for spontaneous emission, $A_{u \ell} = 2.01 \times 10^8$\,s$^{-1}$ \citep{NIST_ASD}, in the solar spectrum the \SrILA{} line shows a weak absorption profile \citep[e.g.,][]{malherbe2007} with an equivalent width of 36\,m{\AA} \citep[][]{moore1966}, consisting of a narrow Doppler core and no wings.
As pointed out by \citet{Faurobert-Scholl1993}, the reason for this is the low abundance of strontium in the solar atmosphere, given by \citep[see][]{asplund2009,bergemann2012}
\begin{equation*}
    A_\mathrm{Sr} = \log_{10}\!\left({N_\mathrm{Sr}}/{N_\mathrm{H}}\right) + 12 = 2.9, 
\end{equation*}
with $N_\mathrm{Sr}$ and $N_\mathrm{H}$ the number densities of strontium and hydrogen, respectively, and the fact that neutral strontium is a minority species.\footnote{According to non-LTE RT calculations in 1D semi-empirical atmospheric models, the ratio between the number densities of neutral and singly ionized strontium
in the photosphere is in the order of $10^{-3}$ and it further decreases with height.}

\begin{figure}[t!]
    \includegraphics[width=0.49\textwidth]{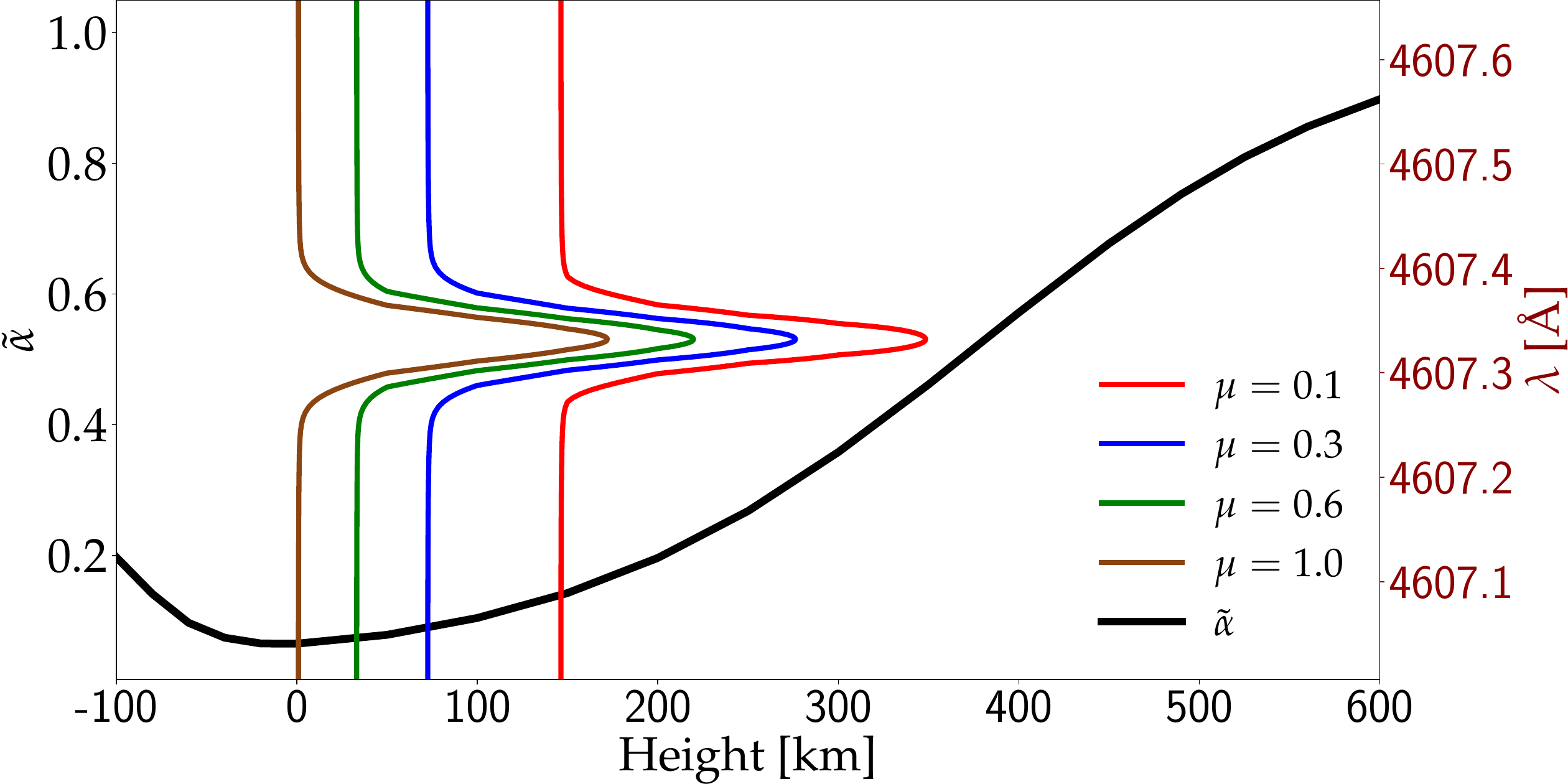}
    \caption{
        \small{Black line: \SrILA{} coherence fraction $\tilde {\alpha}$ (see Eq.~\eqref{eq:coherence_fract}, left axis) as a function of height in the FAL-C atmospheric model.
        Colored iso-lines: heights at which the optical depth is unity as a function of wavelength (right axis), for \mbox{$\mu\in\{0.1,0.3,0.6,1\}$}.
        }
    }
    \label{fig:tauSrI}
\end{figure}

Figure~\ref{fig:tauSrI} shows the formation height of \SrILA{} in model\,C of \citet[][hereafter, FAL-C]{fontenla1993} for different lines-of-sight (LOSs), as calculated with RH \citep[][]{uitenbroek2001multilevel}, together with the coherence fraction, defined as 
\begin{equation}\label{eq:coherence_fract}
\tilde{\alpha} = \frac{\Gamma_R + \Gamma_I}{\Gamma_R + \Gamma_I + \Gamma_E},
\end{equation}
with $\Gamma_R$, $\Gamma_I$, and $\Gamma_E$ the damping constants due to radiative decays, inelastic (de-exciting) collisions, and elastic collisions, respectively.
The quantity $\tilde\alpha$ gives a measure of the relative contribution of scattering processes that are coherent and totally incoherent (i.e., uncorrelated) in frequency in the atomic rest frame: a value close to unity means that the former dominate, while a value close to zero means that the latter dominate.
As is clear from Eq.~\eqref{eq:coherence_fract}, the balance between such processes is determined by the efficiency of elastic collisions (i.e., by the probability that the atom suffers elastic collisions before it de-excites, either radiatively or via inelastic collisions).
From Fig.~\ref{fig:tauSrI} we see that the core of \SrILA{} forms in the photosphere, between approximately 150\,km for a LOS with $\mu=1$ (i.e., along the vertical) and 350\,km for a LOS with $\mu=0.1$, with $\mu$ the cosine of the heliocentric angle.\footnote{These values are in agreement with those obtained from 3D RT calculations  \citep[e.g.,][]{Shchukina2011}.}
{At such heights, the contribution of scattering processes that are coherent in frequency in the atomic rest frame is relevant,
with $\tilde{\alpha}$ ranging between approximately 0.2 and 0.4.
Considering that scattering polarization is very sensitive to the angular and frequency coupling of the incoming and outgoing radiation in scattering processes, the relatively high value of $\tilde{\alpha}$
makes it worth systematically verifying the validity of the CRD approximation for modeling the magnetic sensitivity of \SrILA{}.\footnote{The CRD limit is considered a reasonable approximation because the \SrILA{} line has no wings and the thermal motions of the atoms efficiently redistribute scattered photons in frequency in the line core. 
}}

\subsection{Radiative transfer problem}
\label{sect:RTproblem}

The intensity and polarization of a radiative beam
are fully described by
the Stokes vector $\vec{I}=(I,Q,U,V)^{T}$, where $I$ denotes the intensity, $Q$ and $U$ the linear polarization, and $V$ the circular polarization. 
The propagation of a radiative beam of frequency $\nu\in\mathbb{R}^+$ along a direction $\vec{\Omega}$, specified by the polar angles $(\theta,\chi)\in[0, \pi]\times[0, 2\pi)$, at a spatial point $\vec{r} \in D\subset \mathbb{R}^3$ in the solar plasma is described by:
\begin{equation}
    \vec{\nabla}_{{\vec{\Omega}}} \vec{I}(\vec{r},\vec{\Omega},\nu) = 
   - K(\vec{r},\vec{\Omega},\nu) \vec{I}(\vec{r},\vec{\Omega},\nu) + 
   \bm{\varepsilon}(\vec{r},\vec{\Omega},\nu) ,
   \label{eq:rte}
\end{equation}
where $\vec{\nabla}_{\vec{\Omega}}$ $= \vec{\Omega} \cdot \nabla$ denotes the directional derivative,
$K \in \mathbb{R}^{4 \times 4}$ is the propagation matrix,
and  
$\bm{\varepsilon}(\vec{r},\vec{\Omega},\nu)$ the emission vector \citep[e.g.,][]{degl2006polarization}.

The RT coefficients $K$ and $\bm{\varepsilon}$ depend on the state of the atom, which in turn depends on the radiation field $\vec{I}$ and on the other particles present in the plasma (through collisional processes) via the statistical equilibrium (SE) equations.
The non-LTE RT problem consists of finding a self-consistent solution of the RT equation~\eqref{eq:rte} and the SE equations for the considered atomic system.
The non-LTE RT problem is in general nonlocal, nonlinear, and densely coupled in angular and spectral dimensions.

We formulate the problem in a right-handed Cartesian reference system, with the $z$-axis directed along the vertical. 
The inclination $\theta$ is measured with respect to the positive $z$-axis, while the azimuth is measured in the $xy$-plane, counter-clockwise (for an observer at positive $z$) from the $x$-axis.
The reference direction for positive Stokes $Q$ is taken parallel to the $xy$-plane.

\subsection{Emission vector and propagation matrix}

We consider a two-level atom with an unpolarized and infinitely-sharp lower level (see Sect.~\ref{sec:SrIline}), and we work within the framework of the redistribution matrix formalism (applicable when the SE equations have a closed analytic solution).
Within this formalism, the emission vector is given by the sum of two terms, describing the contributions from scattering (label `sc') and thermal (label `th') processes:
\begin{equation}
	\label{eq:emission_sc_th}
	\vec{\varepsilon}(\vec{r},\vec{\Omega},\nu) =
    \vec{\varepsilon}^{\mathrm{sc}}(\vec{r},\vec{\Omega},\nu) +
	\vec{\varepsilon}^{\mathrm{th}}(\vec{r},\vec{\Omega},\nu) .
\end{equation}
The scattering term is given by
\begin{equation}
    \vec{\varepsilon}^{\mathrm{sc}}(\vec{r},\vec{\Omega},\nu) = k_L(\vec{r})\!\!
	\int_{\mathbb{R}_{+}} \!\!\!\! \mathrm{d} \nu'\! \oint 
	\frac{\mathrm{d} \vec{\Omega}'}{4 \pi} R(\vec{r},\vec{\Omega},\vec{\Omega}',\nu,\nu' ) \vec{I}(\vec{r},\vec{\Omega}',\nu') ,
	\label{eq:scat_int}
\end{equation}
where $k_L(\vec{r})$ is the frequency-integrated absorption coefficient and $R$ is the redistribution matrix, which encodes the SE equations and the physics of scattering (see Sect.~\ref{sec:R_matrix}).
The primed and unprimed variables refer to the incident and scattered radiation, respectively.
The integral in the right-hand side of Eq.~\eqref{eq:scat_int} is generally referred to as the scattering integral.
The explicit expressions of the thermal emissivity $\vec{\varepsilon}^{\mathrm{th}}$ and the propagation matrix $K$ for the considered atomic model can be found in \citet{riva2023RIII}.

The magnetic field enters the redistribution matrix $R$ and the propagation matrix $K$. 
In this work, we account for both the Hanle and Zeeman effects produced by magnetic fields of arbitrary strength (Hanle-Zeeman regime). 
The Hanle effect is encoded in $R$, while the Zeeman effect (i.e., the splitting of the magnetic sublevels) is encoded in both $R$ and $K$.
Continuum processes are included in this work, adding their contributions to the expressions of $K$ and $\varepsilon$ \citep[e.g.,][]{benedusi2022}.

\subsection{Redistribution matrix}
\label{sec:R_matrix}

We consider the redistribution matrix for a two-level atom with unpolarized and infinitely-sharp lower level, in the presence of arbitrary magnetic fields, as derived by \citet{bommier1997masterI,bommier1997masterII}.
This is given by the linear combination of two terms
\begin{equation*}
	R(\vec{r},\vec{\Omega},\vec{\Omega}',\nu,\nu') = 
	R^{\scriptscriptstyle \mathrm{II}}(\vec{r},\vec{\Omega},\vec{\Omega}',\nu,\nu')+ 
	R^{\scriptscriptstyle \mathrm{III}}(\vec{r},\vec{\Omega},\vec{\Omega}',\nu,\nu') ,
\end{equation*}
where $\RII$ and $\RIII$ describe scattering processes that are coherent and totally uncorrelated in frequency, respectively, in the atomic rest frame.
Their relative contribution is set by the coherence fraction $\tilde{\alpha}$, defined in Eq.~\eqref{eq:coherence_fract}.

In this study, we consider and compare four different forms of the redistribution matrix.
In the most general case, 
both $\RII$ and $\RIII$ are treated in their exact AD formulation, namely,
\begin{equation}
R = \RII + \RIII .
\label{eq:Rexact}
\end{equation}
In this form, both $\RII$ and $\RIII$ present a very complex
coupling between all frequencies and propagation directions of the incident and scattered radiation, entailing 
a very high, often prohibitive, computational cost, especially for $\RIII$ \citep[e.g.,][]{riva2023RIII}.
Various approximations have been proposed to loosen such coupling, thereby simplifying the evaluation of $R$ and $\bm{\varepsilon}^{\mathrm{sc}}$.
A widely used one
is to replace the computationally expensive $\RIII$ in its exact form with the limit of CRD
\citep[e.g.,][]{ballester2017transfer}, namely,
\begin{equation}
R = \RII + \RIIICRD .
\label{eq:PRDAD}
\end{equation}
Notably, the frequency and angular dependencies of the incident and scattered radiation are completely decoupled in $\RIIICRD$.
On top of this, one can further apply the well-known AA approximation to $\RII$ \citep[e.g.,][]{rees1982}, namely,
\begin{equation}
R = \RIIAA + \RIIICRD .
\label{eq:PRDAA}
\end{equation}
In $\RIIAA$, only the frequencies of the incident and scattered radiation are still coupled.
Finally, one can consider the limit of CRD, in which radiation scattering is simply modeled as a temporal succession of uncorrelated absorption and emission processes \citep[e.g.,][]{degl2006polarization}, namely
\begin{equation}
    R = \RCRD,
\label{eq:CRD}
\end{equation}
whereby the incident and scattered frequencies and directions are fully decoupled.
The explicit expressions of all the aforementioned redistribution matrices, in both their exact form and under the various approximations, can be found for instance in \citet{riva2023RIII}.
 Hereafter, the scattering descriptions corresponding to the redistribution matrices of Eqs.~\eqref{eq:PRDAD}, \eqref{eq:PRDAA}, and \eqref{eq:CRD} will be referred to as PRD, PRD-AA, and CRD, respectively.
The case of Eq.~\eqref{eq:Rexact} will be referred to as PRD-exact.

\newcommand{\NIST}{\footnote{\href{https://physics.nist.gov/PhysRefData/ASD/lines\_form.html}{https://physics.nist.gov/PhysRefData/ASD/lines\_form.html}}}

\FloatBarrier

\section{Setting and methods}
\label{sec:methods}

We now detail the atmospheric model, the adopted numerical parameters, and the solution methods employed to solve the non-LTE RT problem for polarized radiation for the \SrILA{} line described in Sect.~\ref{sec:rt_problem}.

\subsection{Atmospheric model and spectral and angular grids}
\label{sec:atmo_grids}
We considered a subset of the 1D FAL-C atmospheric model,
consisting of 30 discrete heights ranging from -100\,km to 1378\,km. 
The exclusion of the upper part of the original model is justified by the fact that the \SrILA{} line forms below 500\,km for a LOS close to the limb (see Fig.~\ref{fig:tauSrI}).
We adopted the microturbulent velocities as determined in \citet{fontenla1991}.
Finally, we assumed that no radiation enters the domain through the top boundary, while the bottom boundary is illuminated by an isotropic, spectrally flat, and unpolarized Planckian incident radiation.
The spectral and angular discretizations of the problem consist of $N_{\nu}=131$ frequency points and an angular grid based on the spherical Cartesian rule with $ N_{\theta} = 12$ Gauss-Legendre points for inclination and $N_{\chi}=8$ evenly spaced points for the azimuth.

\referee{
In this work, we focus on assessing the suitability of the CRD assumption in modeling the magnetic sensitivity of the scattering polarization signal of \SrILA{} in a static atmosphere.
We recall that bulk velocity gradients can heavily affect scattering polarization signals, generally enhancing their amplitudes and inducing asymmetries in the Stokes profiles \citep[e.g.,][]{carlin2012scattering}.
The impact of velocities might be even more important when PRD effects, which introduce a strong coupling between the angular and frequency dependencies of the incident and scattered radiation, are relevant \citep[e.g.,][]{guerreiro2023Vel}.
An example of the effects of a height-dependent vertical bulk velocity on the scattering polarization signal of Sr~{\sc i} 4607\,{\AA}, modeled including PRD effects, is shown in Fig.~17 of  \citet{riva2023RIII}.
}

\subsection{Magnetic field configurations}
We consider two magnetic field configurations, motivated by previous observational and theoretical works.
The first configuration is a height-independent micro-structured and isotropic (MSI) magnetic field with a given strength.
Following the results of \citet{trujillo2004}, we consider field strengths in the range of 50--130\,G.
The impact of a MSI magnetic field is modeled by averaging both $K$ and $R$ over an isotropic distribution of magnetic fields of a given strength \citep[e.g.,][]{degl2006polarization}.
As such, the MSI configuration is an idealized representation of small-scale unresolved magnetic fields with randomly distributed orientations on spatial scales smaller than the photon mean free path.\footnote{In the FAL-C model, at a height of about 250\,km, we find a photon mean free path of approximately 300\,km at the line-center wavelength of \SrILA{}.}

Such magnetic fields may be generated by convective and turbulent motions that continuously stretch, twist, and advect magnetic flux, as suggested by small-scale dynamo action \citep[see review by][]{Rempel2023}. 
However, the orientation and strength distribution of these unresolved quiet-Sun photospheric magnetic fields remain debated \citep[e.g.,][]{Bellot2019}.
We note that a MSI magnetic field does not break the axial symmetry of 1D atmospheric models and therefore $U/I$ remains zero. 
Since the Zeeman effect is blind to such fields, also $V/I$ is zero.
The MSI magnetic field affects the emergent polarization exclusively through the Hanle effect, leading to depolarization of the $Q/I$ scattering polarization signal.

The second considered configuration is a magnetic field with a given strength and orientation {(specified by the inclination $\theta_B$ and azimuth $\chi_B$, defined as explained in Sect.~\ref{sect:RTproblem})}, constant throughout the model atmosphere.
{Such magnetic fields, hereafter referred to as deterministic, appear of interest because of the recent observations by \citet{zeuner2022}, which showed a $U/I$ signal compatible with the Hanle rotation produced by a spatially-structured component of the small-scale magnetic field of the quiet solar photosphere}.
For the considered strengths in the range 10--100\,G, these magnetic fields can produce $V/I$ circular polarization signals via the Zeeman effect.
Moreover, they can modify the amplitude of scattering polarization and rotate its plane of linear
polarization, leading to non-zero $U/I$, via the Hanle effect.

\subsection{Solution strategy}
\label{sec:formalsolv}

We solved the non-LTE RT problem for polarized radiation following the solution strategy presented by \citet{benedusi2022,benedusi3Drt2023}.
In this approach, the population of the lower level is a fixed input parameter of the problem.
The population of the Sr~{\sc i} ground level was thus precomputed using the RH code \citep{uitenbroek2001multilevel}, which solved the non-LTE RT problem for unpolarized radiation, considering
a strontium atomic model consisting of 33 levels of Sr~{\sc i} and
the ground level of Sr~{\sc ii}. 
These preliminary calculations also provided the rates for elastic and inelastic collisions, together with the continuum quantities.

We solved the non-LTE RT problem for polarized radiation applying the 1D module of the TRIP code.\footnote{\url{https://github.com/pietrobe/TRIP/}}
For these calculations, we performed the formal solution of the RT equation using the DELO-linear method \cite[e.g.,][]{janettI}.
As iterative solver, we applied FGMRES, accelerated via a tailored physics-based preconditioner \citep[][]{benedusi2021numerical,janett2024numerical}, with a tolerance of $10^{-9}$ as a stopping criterion for the relative residual.

\FloatBarrier
\section{Results}\label{sec:results}

We compare {the emergent Stokes profiles of \SrILA{}, calculated considering} CRD (see Eq.~\eqref{eq:CRD}), PRD-AA (see Eq.~\eqref{eq:PRDAA}), and PRD (see Eq.~\eqref{eq:PRDAD}) scattering descriptions, {first} 
in the presence of MSI magnetic fields and, subsequently, in the presence of deterministic ones.
For completeness, Appendix~\ref{subsec:prdexact} further generalizes this analysis by considering the PRD-exact case, in which both $\RII$ and $\RIII$ are considered in their full angle-dependent form (see Eq.~\eqref{eq:Rexact}).
We computed the emergent radiation for 26 discrete LOSs
with inclination $\mu = \cos{\theta} \in (0,1]$ and azimuth $\chi=0$.
\begin{figure*}[htb]
    \centering
    \includegraphics[width=\textwidth]{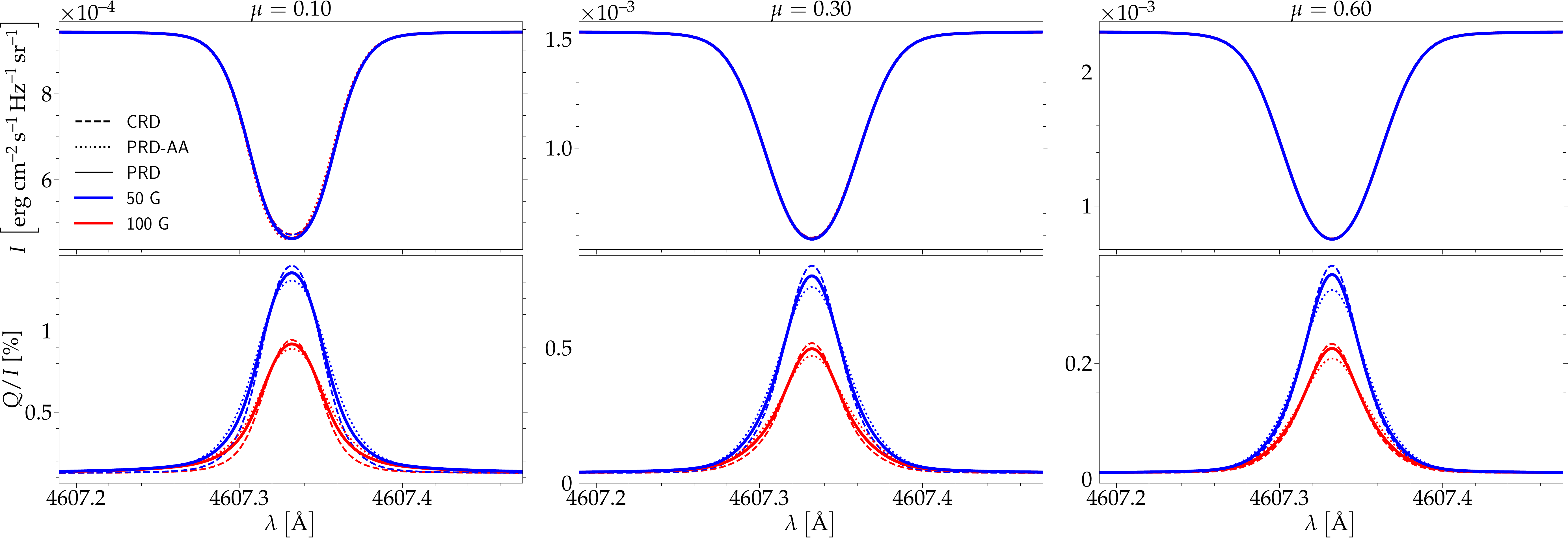}
    \caption{\small{
    Emergent intensity (upper panel) and $Q/I$ (lower panel) \SrILA{} profiles calculated in the FAL-C atmospheric model for \mbox{$\mu\in\{0.1,0.3,0.6\}$} (different columns), including a height-independent MSI magnetic field of 50\,G (blue lines) and 100\,G (red lines), considering CRD (dashed), PRD-AA (dotted), and PRD (solid) scattering descriptions.
    In this setting, Stokes $U/I$ and $V/I$ vanish and are thus not shown. 
    }}

    \label{fig:MSI_profiles}
\end{figure*}
\begin{figure}[htb!]
    \centering
    \includegraphics[width=0.5\textwidth]{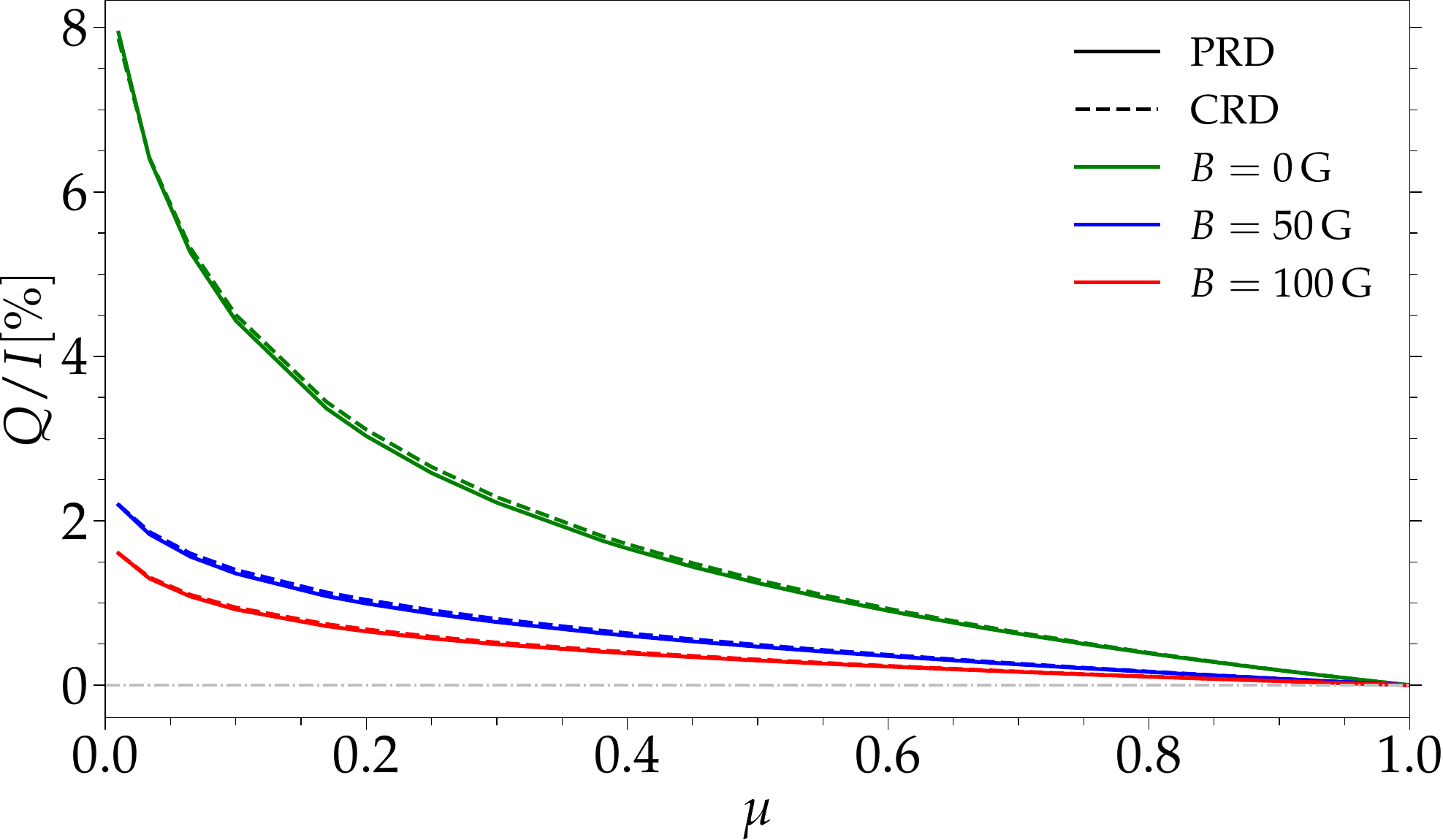}
    \caption{\small{
    \SrILA{} line-center $Q/I$ signal as a function of $\mu$, calculated in the FAL-C atmospheric model including MSI magnetic fields of different strengths, considering the PRD (solid), CRD (dashed) scattering descriptions.
    }}
    \label{fig:cases_MSI_B_CLV_CRD}
\end{figure}

\subsection{MSI magnetic fields}
\label{sec:MSIB}
Figure~\ref{fig:MSI_profiles} presents the $I$ and $Q/I$ profiles computed in the presence of MSI magnetic fields of 50\,G and 100\,G for the CRD, PRD-AA, and PRD scattering descriptions, for three different LOSs.
The discrepancy in $Q/I$ between CRD and PRD results is fairly small.
The agreement between PRD-AA and PRD results is also good, but the relative discrepancy increases moving towards the disk center.
For conciseness, in the following we will only focus on the comparison between CRD and PRD cases, while more details on the error introduced by the PRD-AA approximation are provided in Appendix~\ref{sec:errorAACRD}.
For a broader overview, Fig.~\ref{fig:cases_MSI_B_CLV_CRD} shows the center-to-limb variation of the line-center $Q/I$ signal, confirming the good agreement between CRD and PRD results across all considered LOSs.

\begin{figure}[htb!]
    \includegraphics[width=0.5\textwidth]{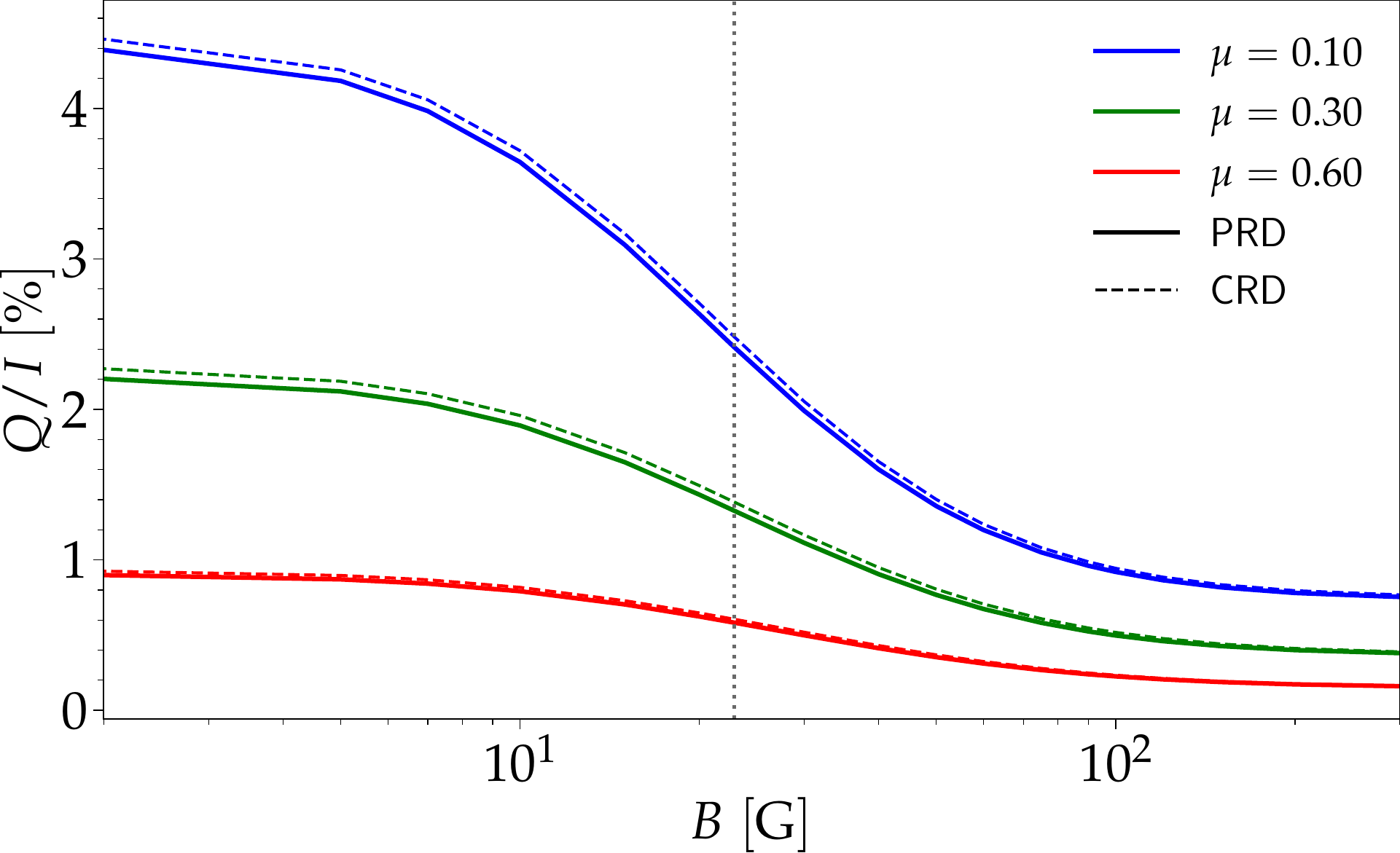}
    \caption{\small{
    \SrILA{} line-center $Q/I$ signal as a function of the strength $B \in [0\,G, 300\,G]$ of a MSI magnetic field, calculated in the FAL-C atmospheric model for $\mu\in\{0.1,0.3,0.6\}$.
    Solid and dashed lines correspond to the PRD and CRD scattering descriptions, respectively.
    The dotted vertical line shows the Hanle critical field $B_H=22.9$\,G.
    }}
    \label{fig:QI_vs_B}    
\end{figure}

Fig.~\ref{fig:QI_vs_B} shows the line-center $Q/I$ signal as a function of the magnetic field strength for CRD and PRD scattering descriptions at different LOSs.
We first note that the ratio of $Q/I$ between the 300\,G case (saturated Hanle regime) and the 0\,G case (no Hanle effect) is close to $1/5$, as theoretically expected 
in the presence of an MSI magnetic field \citep[e.g., Eq.~5.173 of][]{degl2006polarization}.
Although yielding slightly different absolute values, the CRD and PRD calculations exhibit the same trend with the field strength, indicating that they predict the same Hanle magnetic sensitivity.

 As a further analysis, Table~\ref{tab:sensitivityQIcol} reports, for the CRD and PRD scattering descriptions, the sensitivity of the line-center $Q/I$ signal to a perturbation of the MSI magnetic field strength of $\Delta B = 7$\,G, namely,
\begin{equation}
\label{eq:hanle_CRD_PRD}
\Delta(Q/I)(B,\mu) = |Q/I(B,\mu) - Q/I(B+\Delta B,\mu)|.
\end{equation}
The value $\Delta B = 7$\,G corresponds to the solar-cycle variations in quiet-Sun magnetism predicted by \citet{Rempel2020}.
Interestingly, the CRD and PRD calculations agree remarkably well for all the considered cases, with absolute discrepancies below $10^{-2}\%$.
This result has important implications, as it indicates
that the Hanle magnetic sensitivity of the CRD and PRD scattering descriptions is effectively equivalent for practical purposes.
\begin{table}[ht]
\centering
\setlength{\tabcolsep}{10pt}
\renewcommand{\arraystretch}{1.5}
\begin{tabular}{c|c|c|c}
$\mu$ & $B$ [G] & $\Delta(Q/I)_{\mathrm{PRD}}$ [\%] & $\Delta(Q/I)_{\mathrm{CRD}}$ [\%] \\
\hline
\multirow{4}{*}{0.1} & 50 & $1.1 \cdot 10^{-1}$ & $1.2 \cdot 10^{-1}$ \\
 & 75 & $4.2 \cdot 10^{-2}$ & $4.4 \cdot 10^{-2}$ \\
 & 100 & $1.9 \cdot 10^{-2}$ & $2.0 \cdot 10^{-2}$ \\
 & 130 & $1.1 \cdot 10^{-2}$ & $1.2 \cdot 10^{-2}$ \\
\hline
\multirow{4}{*}{0.3} & 50 & $6.6 \cdot 10^{-2}$ & $6.9 \cdot 10^{-2}$ \\
 & 75 & $2.7 \cdot 10^{-2}$ & $2.9 \cdot 10^{-2}$ \\
 & 100 & $1.3 \cdot 10^{-2}$ & $1.4 \cdot 10^{-2}$ \\
 & 130 & $7.6 \cdot 10^{-3}$ & $8.4 \cdot 10^{-3}$ \\
\hline
\multirow{4}{*}{0.6} & 50 & $3.0 \cdot 10^{-2}$ & $3.1 \cdot 10^{-2}$ \\
 & 75 & $1.3 \cdot 10^{-2}$ & $1.4 \cdot 10^{-2}$ \\
 & 100 & $6.9 \cdot 10^{-3}$ & $7.4 \cdot 10^{-3}$ \\
 & 130 & $4.2 \cdot 10^{-3}$ & $4.6 \cdot 10^{-3}$ \\
\end{tabular}
\caption{\small{
Variation of $Q/I$ (see Eq.~\eqref{eq:hanle_CRD_PRD}) corresponding to a change of 7\,G in the strength of a MSI magnetic field. 
The variation is calculated in the FAL-C atmospherc model, for different LOSs and magnetic field strengths, considering PRD and CRD scattering descriptions.
}}
\label{tab:sensitivityQIcol}
\end{table}

\subsection{Deterministic magnetic fields}
\label{sec:detB}

\begin{figure*}[tbp]
    \centering
    \includegraphics[width=\textwidth]{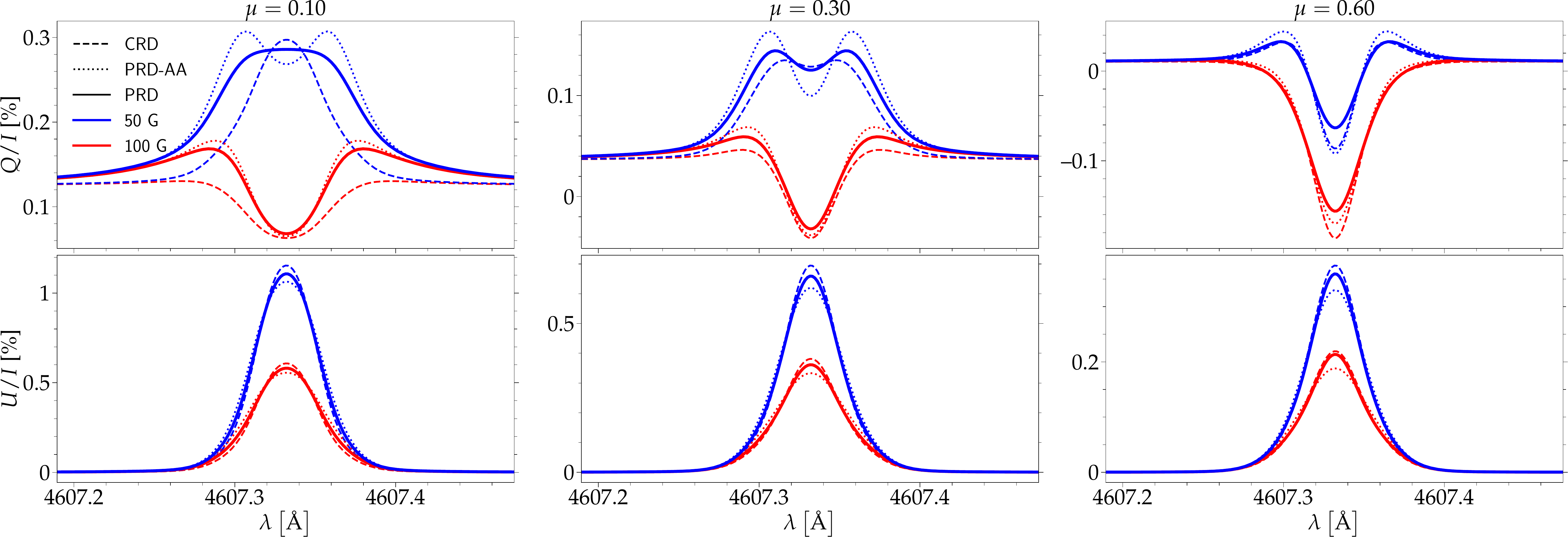}
    \caption{\small{
    Emergent $Q/I$ (upper panels) and $U/I$ (lower panels) \SrILA{} profiles calculated in the FAL-C atmospheric model for $\mu\in\{0.1,0.3,0.6\}$ (different columns) and $\chi=0$, in the presence of height-independent deterministic magnetic fields of 50\,G (blue lines) and 100\,G (red lines) with $\theta_B = \pi/2$ and $\chi_B = 0$, considering CRD (dashed), PRD-AA (dotted), and PRD (solid) scattering descriptions.
    The intensity profiles are not shown, as they are equivalent to those obtained for the MSI magnetic-field configuration shown in Fig.~\ref{fig:MSI_profiles}.
    }}
    \label{fig:Inv_FSHE_profiles}
\end{figure*}

In this section, we assess the impact of the CRD approximation on the modeling of the \SrILA{} scattering polarization in the presence of deterministic magnetic fields.
Figure~\ref{fig:Inv_FSHE_profiles} presents the $Q/I$ and $U/I$ profiles computed in the presence of deterministic magnetic fields of 50\,G and 100\,G, considering the CRD, PRD-AA, and PRD scattering descriptions, for three different LOSs. 
In contrast to the results of Sect.~\ref{sec:MSIB} for MSI magnetic fields, these profiles show that CRD and PRD calculations are not in good agreement in general.
For sake of conciseness, also in this case, the PRD-AA results are not considered in the following analysis.

\begin{figure*}[tbp]
    \centering
    \includegraphics[width=\textwidth]{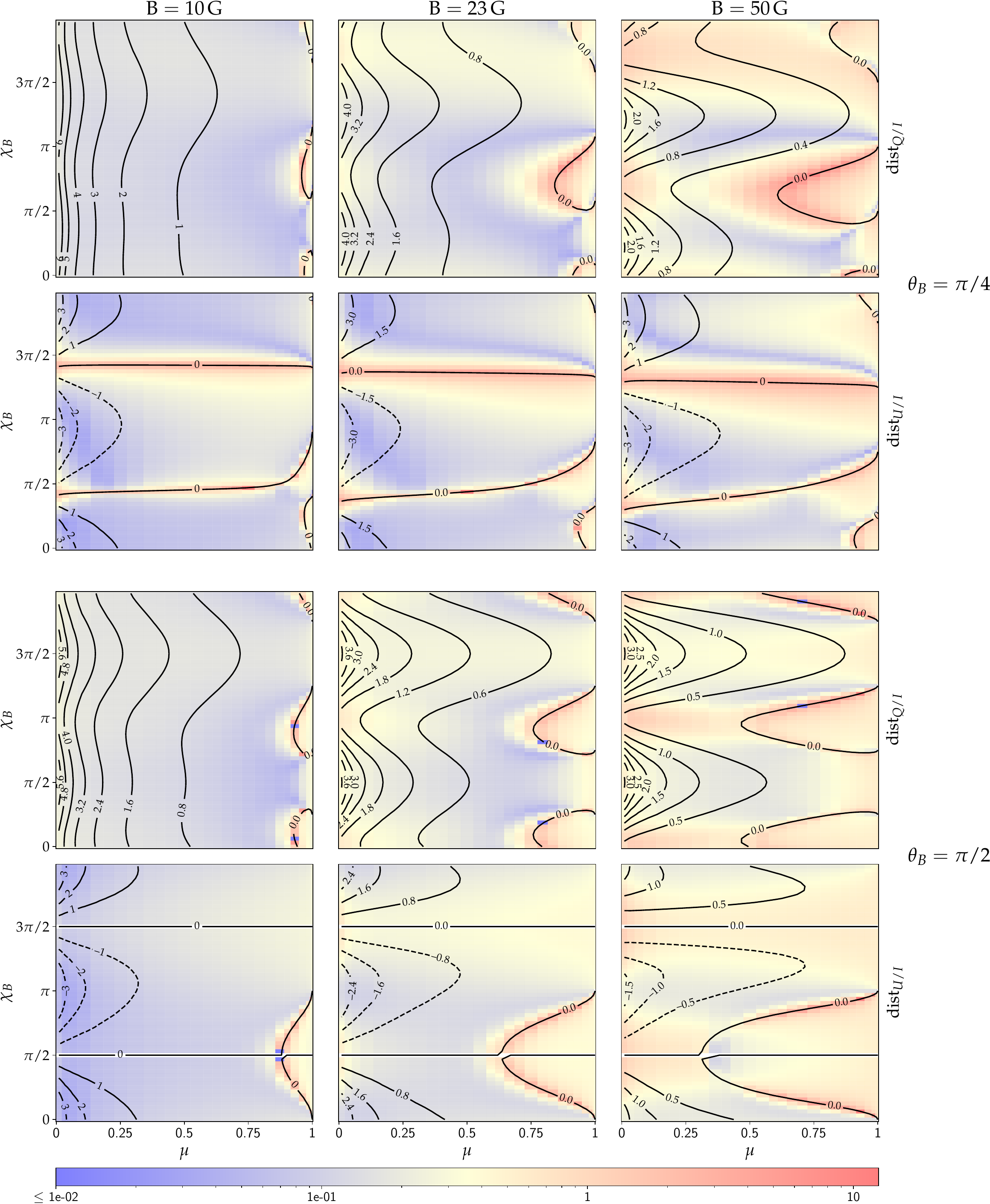}
    \caption{\small{
    2D map of the profile distances between CRD and PRD results ($\mathrm{dist}_X$, with $X \in \{Q/I,\, U/I\}$, see Eq.~\eqref{eq:distPC}), as a function of $\mu$ (horizontal axis) and $\chi_B$ (vertical axis), for magnetic field strengths of 10\,G (first column), 23\,G (second column), and 50\,G (third column). The upper block of panels {corresponds to $\theta_B = \pi/4$},
    the bottom block {to $\theta_B = \pi/2$}.
    Within each block, the first row {shows $\mathrm{dist}_{Q/I}$},
    the second row {$\mathrm{dist}_{U/I}$}.
    In all panels, the LOS has azimuth $\chi=0$.
    The logarithmic color scale encodes the profile distance, while isolines show the line-center polarization amplitude (in \%).
    }}
    \label{fig:dist_QI_UI}
\end{figure*}

{In Fig.~\ref{fig:dist_QI_UI}, we map the discrepancy} between PRD and CRD calculations across a wider range of deterministic magnetic field configurations. 
{Specifically, we consider} a grid of magnetic field strengths ($B=10$\,G, 23\,G, and 50\,G), inclinations ($\theta_B = \pi/4$ and $\pi/2$), and azimuths ($\chi_B \in [0, 2\pi)$).
{To quantify the discrepancy, also considering profile shape differences, we measure}
the Euclidean distance between normalized CRD and PRD profiles, namely:
\begin{equation}
    \mathrm{dist}_X(\mu) = \sqrt{ \sum_{i=1}^{N_\nu} 
    \left[ \frac{X_\mathrm{PRD}(\nu_i, \mu) - X_\mathrm{CRD}(\nu_i, \mu)}{m_X(\mu)} \right]^2 } ,  
    \label{eq:distPC}
\end{equation}
with
\[m_X(\mu) = \max_i \left| X_\mathrm{PRD}(\nu_i, \mu) \right|,\]
where $X \in \{Q/I,\, U/I\}$ denotes the fractional Stokes parameter under consideration, and the sum runs over the $N_\nu$ frequency points. 
The normalization by $m_X(\mu)$, i.e. the maximum PRD amplitude at the given LOS, allows configurations with very different polarization levels to be compared on a common scale. 
This metric captures the combined effect of amplitude and shape differences on the full spectral profile.
Isolines in Fig.~\ref{fig:dist_QI_UI} indicate points with the same line-center polarization amplitude in each configuration.
We note the presence of regions with large distances (red areas in the heatmap), indicating configurations for which the CRD and PRD profiles differ significantly.
These regions are primarily clustered around isolines corresponding to weak polarization signals, close to the disk center ($\mu \gtrsim 0.6$).
For the weak polarization signals in these regions, the contribution of $\RII$ is clearly not negligible (see also Fig.~\ref{fig:tauSrI}), and it is important to consider its exact AD expression, which accounts for the detailed coupling between the frequencies and directions of the incident and scattered radiation.
This result suggests that for these weak signals it might be interesting to also consider $\RIII$ in its exact AD form.
This is analyzed in Appendix~\ref{subsec:prdexact}, where we present some \SrILA{} signals for which a full AD treatment of $\RIII$ has a clear impact on both the amplitude and shape.
\section{Discussion and conclusions}
\label{sec:conclusions}

In this paper, we investigated the impact of considering the limit of CRD to model the scattering polarization of the photospheric \SrILA{} line, as well as its sensitivity to a range of magnetic field configurations relevant for Hanle diagnostics.
We performed non-LTE RT calculations in a semi-empirical 1D plane-parallel atmospheric model, including first MSI magnetic fields and, subsequently, deterministic ones.

In the presence of MSI magnetic fields, the CRD approximation is sufficiently robust, producing scattering polarization signals that are very close to those obtained with the PRD description. Notably, our results indicate that the Hanle magnetic sensitivity derived from CRD and PRD scattering descriptions is effectively equivalent for practical purposes.
However, we do not recommend the use of the PRD-AA scattering description, as it can lead to reduced accuracy in the results.

In the presence of deterministic magnetic fields, the CRD description remains robust, provided that the scattering polarization signals are sufficiently strong (typically exceeding 0.3\%).
Although small, PRD effects can, however, become non-negligible in the case of weak scattering polarization signals, leading to noticeable differences in the emergent profiles relative to the CRD description.
Discrepancies between CRD and PRD calculations are highly directional in nature, emerging only for specific combinations of magnetic field directions and LOSs.
Additionally, our results indicate that when the PRD description appears to be relevant, the full angle-dependent treatment of $\RIII$ could also play an important role.

We therefore confirm that the CRD description
reliably captures the relevant physics to model
the scattering polarization signal of the photospheric \SrILA{} line.
However, we draw attention to the case of weak scattering polarization signals, in the presence of deterministic magnetic fields, for which a PRD description may become necessary.
This quantitative assessment is particularly relevant in light of the unprecedentedly accurate spectropolarimetric observations of the \SrILA{} line that will be provided by the next generation of instruments and large solar telescopes, such as DKIST and EST, which could access the spatial scales at which the small-scale magnetic field of the quiet photosphere, often treated as unresolved, may reveal coherent observable structures.

A further step of our analysis will consist of a careful comparison between CRD and PRD scattering descriptions for the \SrILA{} line in comprehensive 3D atmospheric models\referee{, including bulk velocities,} with the TRIP code \citep{benedusi3Drt2023}.
In \referee{dynamic} 3D scenarios, symmetry-breaking effects arising from \referee{velocity gradients and} horizontal inhomogeneities in the solar plasma may have a significant impact on scattering polarization signals.
Moreover, accurate RT calculations in high-resolution 3D quiet-Sun models  allow  modeling  Hanle depolarization in the \SrILA{} line induced by small-scale photospheric magnetic fields, without the need to rely on the MSI description.

\begin{acknowledgements}
This work was financed by the Swiss National Science Foundation (SNSF) through grant 200021-231308. GJ, FR, and FZ acknowledge the financial support from SNSF through grants CRSK-2\_235805, CRSK-2\_237849, and PZ00P2\_215963. IRSOL is supported by the Swiss Confederation (SEFRI), Canton Ticino, the city of Locarno, and the local municipalities. 
\\
\referee{
We thank the anonymous referee for useful comments that allowed improving the quality of the paper.
}
\\
\textbf{Software}: Matplotlib \citep{matplotlib}, Numpy \citep{numpy}
\end{acknowledgements}

\bibliographystyle{aa}
\bibliography{bibfile}
\appendix
\ifshowlinenumbers\else\nolinenumbers\fi
\section{Impact of CRD approximation for the observer's frame $\RIII$}
\label{subsec:prdexact}

A widely-used lightweight approximation for $\RIII$ consists of evaluating its expression under the assumption of CRD in the observer's frame. 
While the adequacy of this approximation for modeling intensity profiles has been firmly established, its suitability for modeling scattering polarization signals has been investigated only in a limited number of studies. 
In particular, \citet{riva2023RIII} showed that significant differences can arise in weak linear polarization signals of the \SrILA{} line when this approximation is employed.

\begin{figure}[htb]
    \centering
    \includegraphics[width=0.4\textwidth]{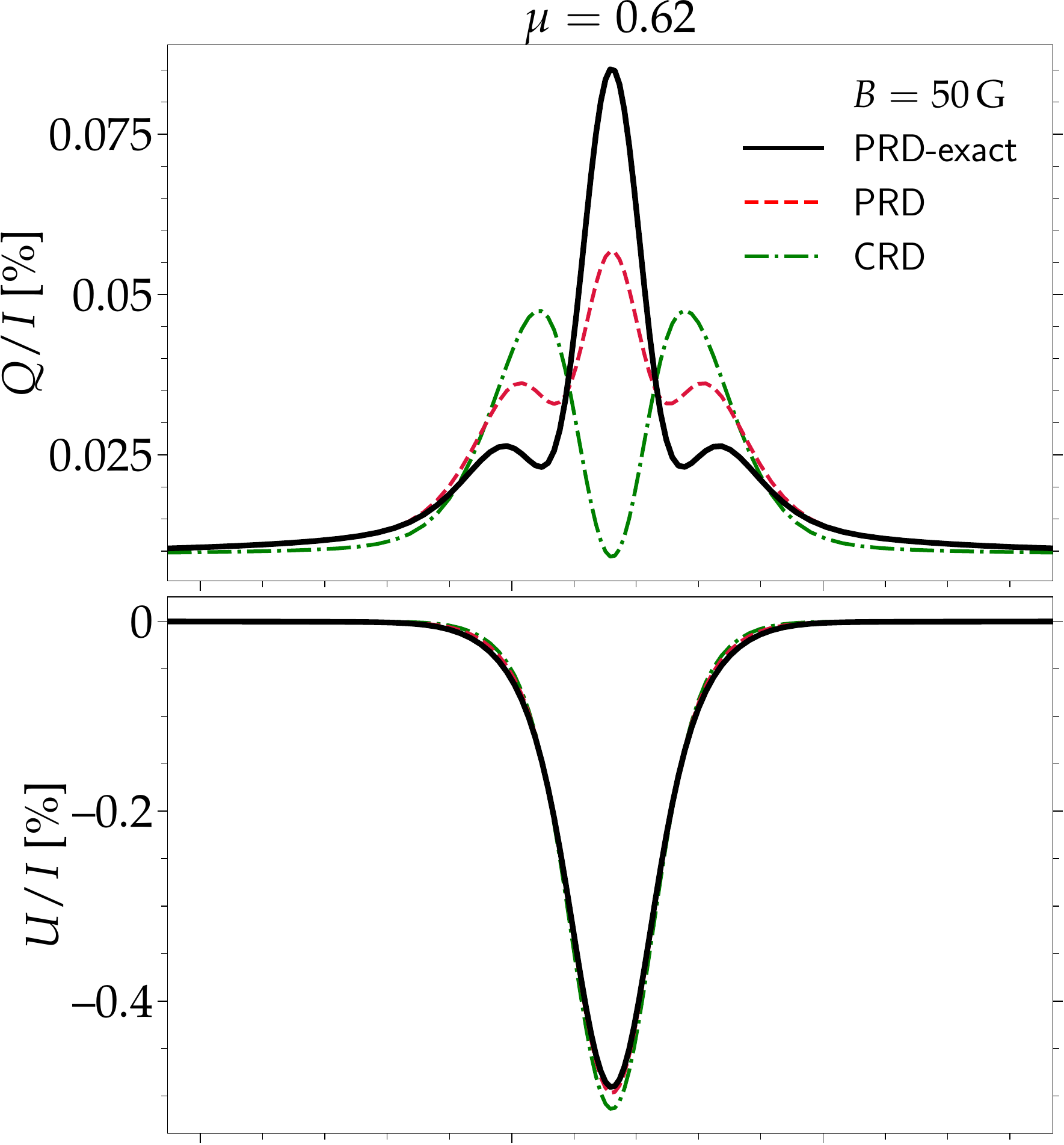}

    \medbreak
    \includegraphics[width=0.4\textwidth]{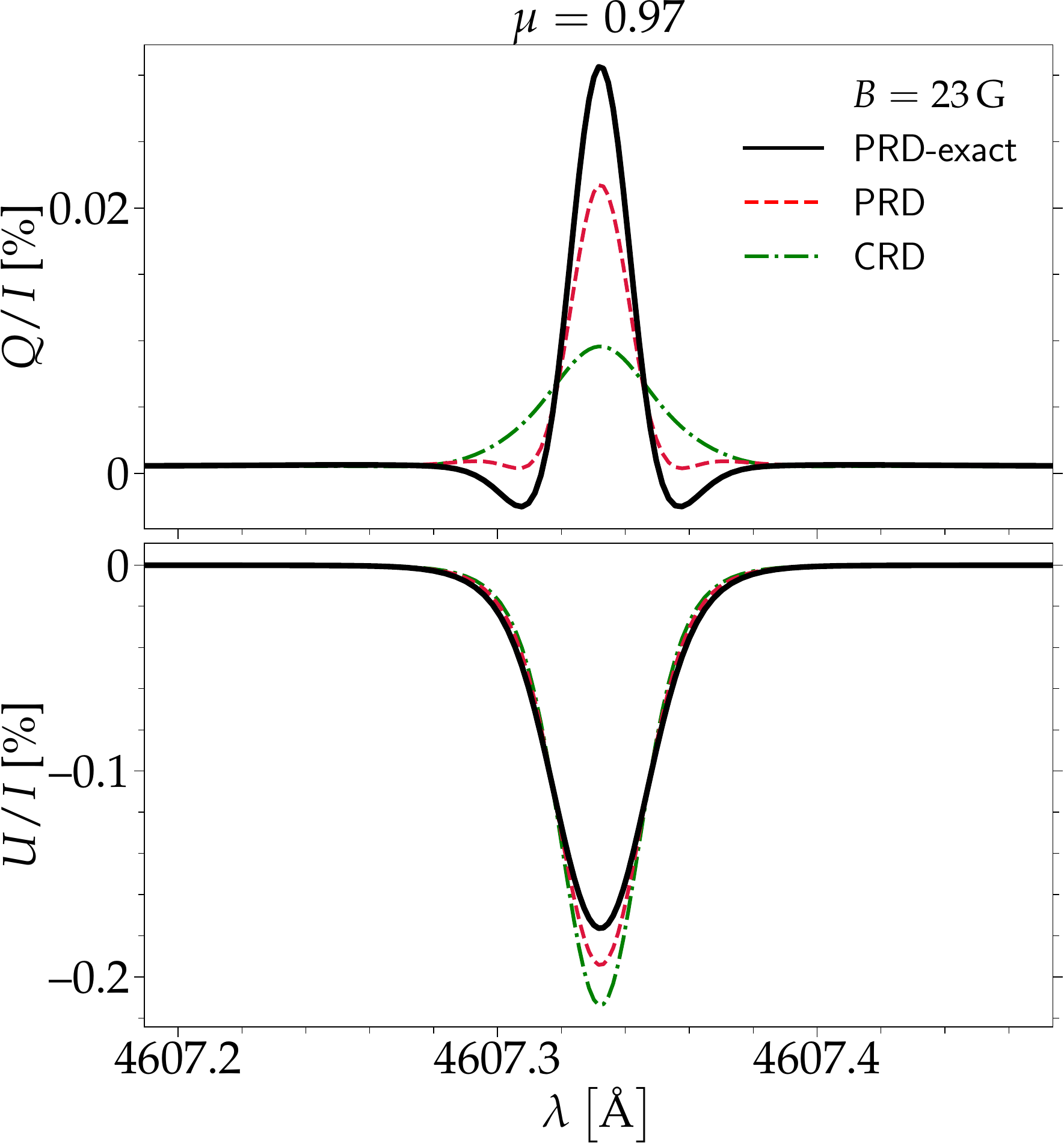}
    
    \caption{\small{
    \emph{Upper panels:} \SrILA{} emergent $Q/I$ and $U/I$ profiles obtained from CRD (dotted-dashed green), PRD (dashed red), and PRD-exact (solid black) calculations in the FAL-C atmospheric model for a LOS with $\mu=0.62$, in the presence of a height-independent deterministic magnetic field with $B=50$\,G, $\theta_B=\pi/4$, and $\chi_B=3\,\pi/4$.
    \emph{Lower panels:} same as upper panels but for a LOS with $\mu = 0.97$ and a magnetic field with $B=23$\,G, $\theta_B = \pi/4$, and $\chi_B=\pi$. 
    }
    }
    
    \label{fig:RIIIexHanle}
\end{figure}

The calculations presented in Sec.~\ref{sec:detB} demonstrate that, in the presence of deterministic magnetic fields, PRD effects can have a significant impact on weak scattering polarization signals of \SrILA{}, leading to noticeable differences in the emergent profiles compared to those obtained under the CRD approximation.
In Fig.~\ref{fig:RIIIexHanle}, we compare the results of PRD-exact calculations to PRD and CRD ones for different LOSs, showing that, for weak scattering polarization signals, the impact of the CRD approximation for the observer's frame $\RIII$ becomes non-negligible.
In particular, the top panel of Fig.~\ref{fig:RIIIexHanle} illustrates a Hanle rotation scenario, where the combined contributions of $\RII$ and the exact $\RIII$ lead to a triple-peak-structure  profile with a more pronounced central peak.
The lower panel of Fig.~\ref{fig:RIIIexHanle} illustrates a forward-scattering Hanle effect  signal.
{ The profile produced by PRD effects in this geometry}, consisting of a sharp central peak flanked by two smaller peaks of opposite sign in the near wings {\citep[see also][]{belluzzi2024accurate}}, is significantly enhanced in PRD-exact calculations.
These two cases indicate that, whenever PRD effects play a significant role in the formation of weak scattering polarization signals, the exact treatment of $\RIII$ can also have a non-negligible impact on the resulting profiles.

\newpage

\section{Error analysis of the CRD and PRD-AA}
\label{sec:errorAACRD}

\begin{figure}
    \centering
    \includegraphics[width=1\linewidth]{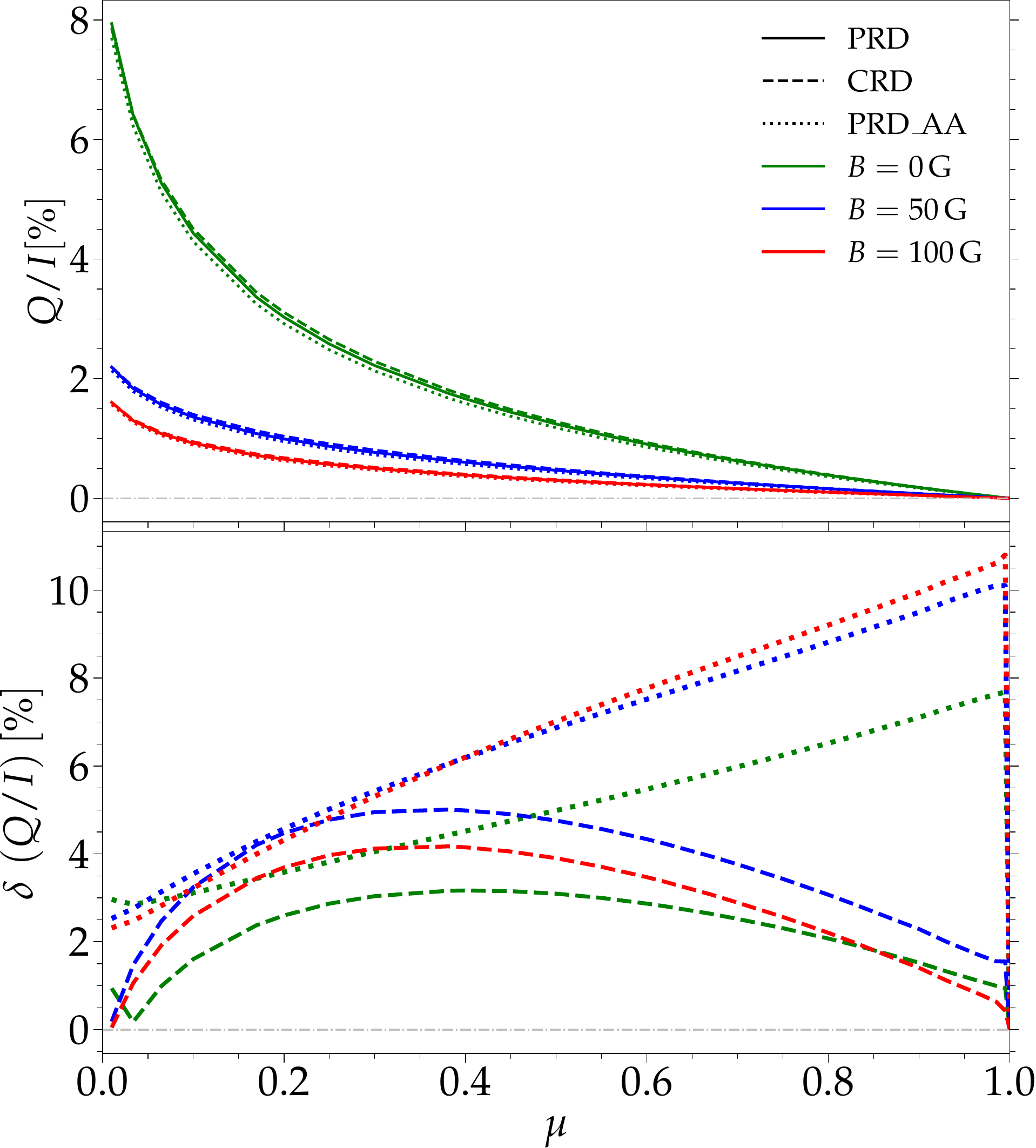}
    \caption{\small{
    \emph{Upper panel}: {\SrILA{} line-center $Q/I$ signal as a function of $\mu$, calculated in the FAL-C atmospheric model including MSI magnetic fields of different strengths, considering} the PRD (solid), CRD (dashed) and PRD-AA (dotted) scattering descriptions. 
    \emph{Lower panel}: relative errors (see Eq.~\eqref{eq:error}) as a function of $\mu$ for CRD (dashed) and PRD-AA (dotted).
    }}
    \label{fig:errorCRDAA}
\end{figure}

Here we present the relative differences between PRD, CRD, and PRD-AA results, assuming an MSI 
magnetic field.
The upper panel of Figure~\ref{fig:errorCRDAA} shows that the 
center-to-limb variations at the central line wavelength 
predicted by the three models are practically indistinguishable. 
However, the relative error reported in the lower panel
\begin{equation}
\delta(Q/I)[\%] = 100 \, \frac{\lvert {\mathrm{PRD}_{\lambda_0}} - {\mathrm{M}_{\lambda_0}} \rvert}{\lvert {\mathrm{PRD}_{\lambda_0}} \rvert} ,
\label{eq:error}
\end{equation}
with $\mathrm{M}\in \{ \mathrm{CRD},\mathrm{PRD}$-$\mathrm{AA} \}$ {and $\lambda_0$ the line-center wavelength},  reveals a systematic difference between the two approximations: the error of PRD-AA grows toward disk center and remains larger than the CRD error across the whole range of $\mu$.
This analysis shows that, at least in the presence of MSI magnetic fields, CRD is a more accurate approximation to the PRD model than PRD-AA.

\end{document}